14 December **2014**

# Observation of subluminal twisted light in vacuum

Frédéric Bouchard[1,2], Jérémie Harris[1,2], Harjaspreet Mand[1,2], Robert W. Boyd[1,2,3], Ebrahim Karimi[1,2]*

[1]Department of Physics, University of Ottawa, 25 Templeton St., Ottawa, Ontario, K1N 6N5 Canada.

[2]The Max Planck Centre for Extreme and Quantum Photonics, University of Ottawa, Ottawa, Ontario, K1N 6N5, Canada.

[3]Institute of Optics, University of Rochester, Rochester, New York, 14627, USA.

*Correspondence to: Ebrahim Karimi (ekarimi@uottawa.ca).

Einstein's theory of relativity establishes the speed of light in vacuum, $c$, as a fundamental constant. However, the speed of light pulses can be altered significantly in dispersive materials. While significant control can be exerted over the speed of light in such media, no experimental demonstration of altered light speeds has hitherto been achieved in vacuum for “twisted” optical beams. We show that “twisted” light pulses exhibit subluminal velocities in vacuum, being slowed by 0.1% relative to $c$. This work does not challenge relativity theory, but experimentally supports a body of theoretical work on the counterintuitive vacuum group velocities of twisted pulses. These results are particularly important given recent interest in applications of twisted light to quantum information, communication and quantum key distribution.

Velocity, the rate at which an object changes its position in time, is well-defined for newtonian particles, but cannot in general be unambiguously assigned to waves. For the specific and unphysical case of a monochromatic plane wave, however, a propagation rate, referred to as the phase velocity, can be attributed to the beam phase-front. The phase velocity $v_{\mathrm{ph}}$ of a monochromatic plane wave in a medium with refractive index $n$ is given by $c/n$. A more general expression for $v_{\mathrm{ph}}$ is required when considering the propagation of a light beam with phase-front $\Phi(\boldsymbol{r})$, $v_{\mathrm{ph}} = \omega/|\nabla\Phi|$, where $\omega$ denotes the angular frequency of the beam and $\nabla$ represents the gradient with respect to the spatial coordinate, $\boldsymbol{r}$ *(1)*. A pulsed beam, which is spatiotemporally localized, is comprised of an infinite superposition of monochromatic waves, each of which propagates at a distinct phase velocity $v_{\mathrm{ph}}(\omega)$. It is the constructive and destructive interference among these frequency components that gives rise to the pulse shape and position. As a result, the pulse propagates at a speed different from that of the individual monochromatic waves of which it is composed. The speed at which the pulse envelope propagates is referred to as the group velocity, $v_{\mathrm{g}}$, and is given by $v_{\mathrm{g}} = |\partial_\omega \nabla\Phi|^{-1}$, where $\partial_\omega$ stands for differentiation with respect to $\omega$ *(1)*, see Supplementary Materials (SM) for more details.

The refractive index of a non-dispersive medium does not depend upon the frequency of light being considered. Consequently, the phase and group velocities associated with a plane wave propagating along a non-dispersive medium’s $z$-axis will take on the values $v_{\mathrm{ph}} = v_{\mathrm{g}} = c/n$, since $\Phi = (\omega\, n/c)z$. By contrast to plane waves, the phase and group velocities of light pulses can differ by orders of magnitude in dispersive media such as cold atomic clouds *(2)*, atomic vapours *(3,4)* and structurally engineered materials *(5-7)*. Under such exotic conditions, pulse group velocities can be rendered greater or smaller than $c$, or even negative *(8)*.

Here we investigate the exotic group velocities exhibited by Laguerre-Gauss (LG) modes in vacuum. In particular, we observe and explain subluminal effects that arise due to the twisted nature of the optical phase front. We use an experimental setup that employs non-linear intensity autocorrelation to measure relative time delays between Gaussian and twisted beams, and show these time delays to be significant, in some cases reaching several tens of femtoseconds.

## The group velocity of twisted light

Despite their mathematical simplicity, plane waves carry infinite energy and therefore are unphysical. More complex waves that can only be approximated even under ideal experimental conditions, such as Bessel beams and evanescent waves, have been studied for their exotic group velocities in vacuum *(9-10)*. A recent publication has also reported slow-light effects in vacuum *(11)* for both Gaussian and Bessel-like beams. The work demonstrated light delays as large as ~ 27 fs. These effects are comparable in magnitude to those reported in our study. Giovannini et al. reported that the optical group delay increased as the square of the diameter of the beam, all other parameters being equal. They interpreted this result in terms of a ray-optics model (which they validated through a wave-optics analysis), in which the slow-light effect occurs because a ray travelling from the edge of the beam to the focus travels a larger distance than an axial ray. This model leads to the prediction that the group delay scales as the square of the beam diameter. In the present work, we report that the group delay increases linearly with the OAM value $\ell$ of the beam. Because the diameter of an LG beam scales as the square root of the $\ell$ value, the scaling law that we observe is consistent with that reported by Giovannini et al. We have interpreted this dependence (see SM) to arise as a consequence of the twisted nature of the optical wavefront. While both models lead to the same scaling law, it is not clear at present that their predictions are identical. Questions related to the mutual compatibility of these approaches remain open, and invite further investigation.

Physically realizable beams, which carry finite energy, possess spatial phase and intensity structures differing from those of plane waves. Laguerre-Gauss modes are among the most commonly encountered examples of such beams, and are solutions to the paraxial wave equation. It may therefore be more transparent to frame the initial theoretical development in the language of pure LG modes, which serve as a more natural basis in which to consider slow light effects in vacuum arising from a twisting of optical wavefronts. Notwithstanding the aesthetic and pedagogical appeal of a pure-LG mode theory, our experiment is carried out using vortex beams, such as HyperGeometric-Gaussian (HyGG), as these are more readily generated experimentally for reasons that will be made clear later *(13)*. LG modes are an orthonormal and complete set, in terms of which any arbitrary paraxial mode can be expanded *(14)*, including HyGG modes. They are characterized by azimuthal and radial mode indices, $\ell$ and $p$, respectively. These modes are eigenstates of orbital angular momentum (OAM), and in vacuum carry OAM values of $\ell\hbar$ per photon along their propagation direction *(15,16)*. LG modal transverse intensity profiles feature intensity maxima at $r_{\max} = w(z)\sqrt{|\ell|/2}$, where $w(z)$ is the beam radius upon propagation *(17)*. The LG mode phase-fronts have helical structures, and in vacuum are given by

$$\Phi^{\mathrm{LG}}(\boldsymbol{r};\omega) = \frac{\omega}{c}z + \frac{\omega\, r^2}{2cR} + \ell\varphi - (2p + |\ell| + 1)\zeta, \tag{1}$$

where $R := R(z,\omega)$ is the radius of curvature of the beam phase-front, $\zeta := \zeta(z,\omega)$ is the Gouy phase, defined in the SM, and $r, \varphi, z$ are the standard cylindrical coordinates. The dependence exhibited by $\Phi^{\mathrm{LG}}$ on its spatial coordinates $\boldsymbol{r}$, angular frequency $\omega$, and the indices $\ell$ and $p$ suggests that the phase velocity of these LG modes also depends on these values. Indeed, when explicitly calculated, the phase velocity is found to depend on $\boldsymbol{r}$, $\omega$, $\ell$ and $p$, i.e. $v_{\mathrm{ph}}(r, z; \omega; p, \ell)$. This concept is illustrated in Fig. 1, which shows the spatial dependence of the phase-fronts associated with Gaussian and spherical waves. This phase velocity leads to the conclusion that $v_{\mathrm{g}}$ is a function of $r, z, \omega, p$ and $\ell$, i.e. $v_{\mathrm{g}}(r, z; \omega; p, \ell)$.

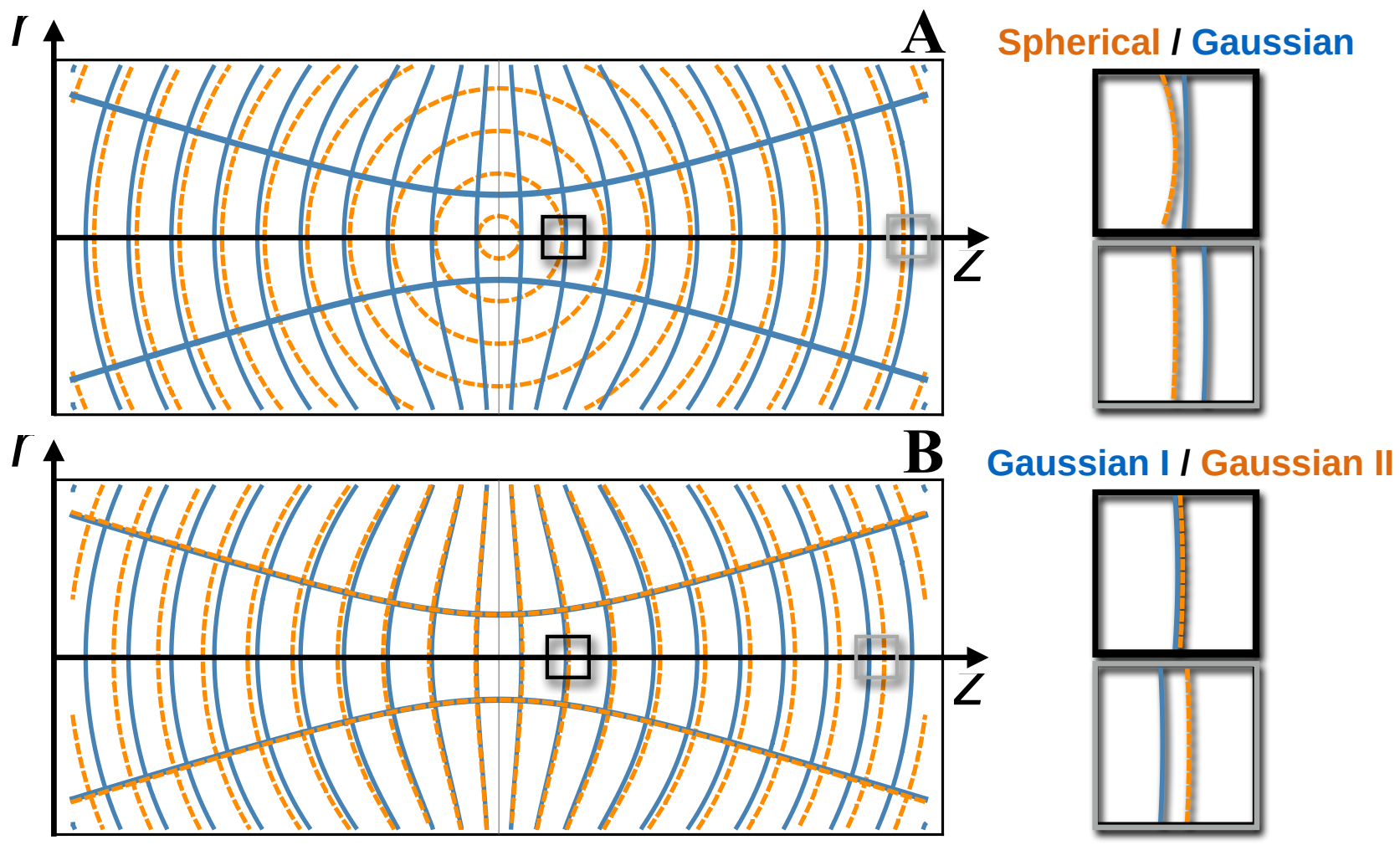

**Fig. 1. Phase distribution of Gaussian and spherical waves upon propagation in vacuum. A** Comparison between spherical (dashed orange) and Gaussian (solid blue) wave phase-fronts having identical wavelengths. As the two beams propagate away from the focus, the phase fronts of the spherical and Gaussian waves begin to separate. This accrued phase discrepancy is associated with the Gouy phase *(12)*. **B** Illustration of the differing wavefront distributions of two Gaussian beams (dashed orange and solid blue, respectively) differing in wavelength by 2%. Insets show enlarged views of the phase-fronts near the beam waist (black colour box) and approximately at 10 wavelengths from the waist (gray colour box). The radial dependence of the difference between the phase-fronts associated with the two beams is apparent in both figures. This can be understood to be a consequence of diffraction; the longer wavelength (orange) beam diverging more rapidly than the shorter wavelength (blue).

The group velocity is calculated by means of a procedure described in the SM, where we also show that the slow-light effect arises largely from the structure of the wavefront radius of curvature, as distinct from the Gouy phase effect already discussed in the theoretical literature *(18-20)*. Indeed, the Gouy phase effect is an order of magnitude smaller than that observed in our experiment. Notably, the slow light effects investigated here are observed to arise due to the twisting of the optical phase front itself, which causes the beam's intensity maximum to follow a hyperbolic trajectory. The group velocities associated with LG modes having different $p$ and $\ell$ indices are shown as functions of distance in Fig. 2. Figure 2-**A** demonstrates that beams with $\ell = 0$ propagate at sub- and super-luminal speeds depending on propagation distance for all values of $p$. Specifically, for propagation distances bounded by $|z| \leq z_R$, these modes exhibit superluminal speeds, which at the waist increase linearly by $\delta v_{\mathrm{g}}^{(p)} = 2.5 \times 10^{-5}\, c\, p$ for the fairly typical case of light at $795\ n$m with a beam waist of $100\ \mu$m. At the positions $z = \pm z_R$, where $z_R = \omega\, w_0^2/(2c)$ is the Rayleigh range, all modes propagate at $c$. Beyond $z_R$, subluminality can be observed. Despite their fast-light behaviour at the waist, beams with $p \neq 0$ still lag behind the Gaussian mode due to their having experienced subluminal velocities prior to $z_R$. The competing slow- and fast-light effects characterizing the propagation of $p \neq 0$ modes cancel to a significant extent for propagation distances beyond $z_R$, rendering $p$-index induced pulse delays extremely difficult to measure. By contrast, for a fixed $p = 0$, LG modes are found to exhibit slow-light behaviour for all propagation distances and all values of $\ell$, as shown in Fig. 2-**B**. In this case, light speed reduction at the waist increases linearly with $\ell$ by $\delta v_{\mathrm{g}}^{(\ell)} = -1.6 \times 10^{-4}\, c\, \ell$ for light at $795\ n$m with a beam waist of $100\ \mu$m. Although this effect is small, the time delay experienced by $\ell \neq 0$ pulses increases monotonically with propagation distance, reaching measurable values far from the waist. Nonetheless, the expected time delays for $\ell \neq 0$ relative to $\ell = 0$ Gaussian pulses are on the order of a few femtoseconds for the wavelength of light and beam waist considered, since the portions of an $\ell \neq 0$ beam that are maximally intense experience more pronounced phase front curvature effects. Therefore, the need for a highly accurate arrival time measurement strategy is clearly indicated *(11,21)*.

## Experimental results

A schematic of our experimental setup is shown in Fig. 3. We measured the relative time delay between a Gaussian reference pulse and an $\mathrm{HyGG}_\ell = \sum_p c_p \mathrm{LG}_{p,\ell}$ pulse by implementing a technique analogous to non-collinear intensity auto-correlation (see the SM). We spatially overlap these non-collinear pulsed beams inside a $\beta$-barium borate (BBO) nonlinear crystal. A time delay is introduced between the two pulses using an optical delay stage. When this delay is minimized, the pulses are spatially and temporally overlapped within the crystal, leading to maximization of the non-collinear second harmonic generation output pulse intensity. In this way, time delays experienced by the test beam can be detected by measuring the delay stage movement required to restore maximal pulse overlap. Using this technique, changes in relative arrival times of the Gaussian and twisted pulses induced by increases in HyGG mode $\ell$ index can be measured within femtoseconds.

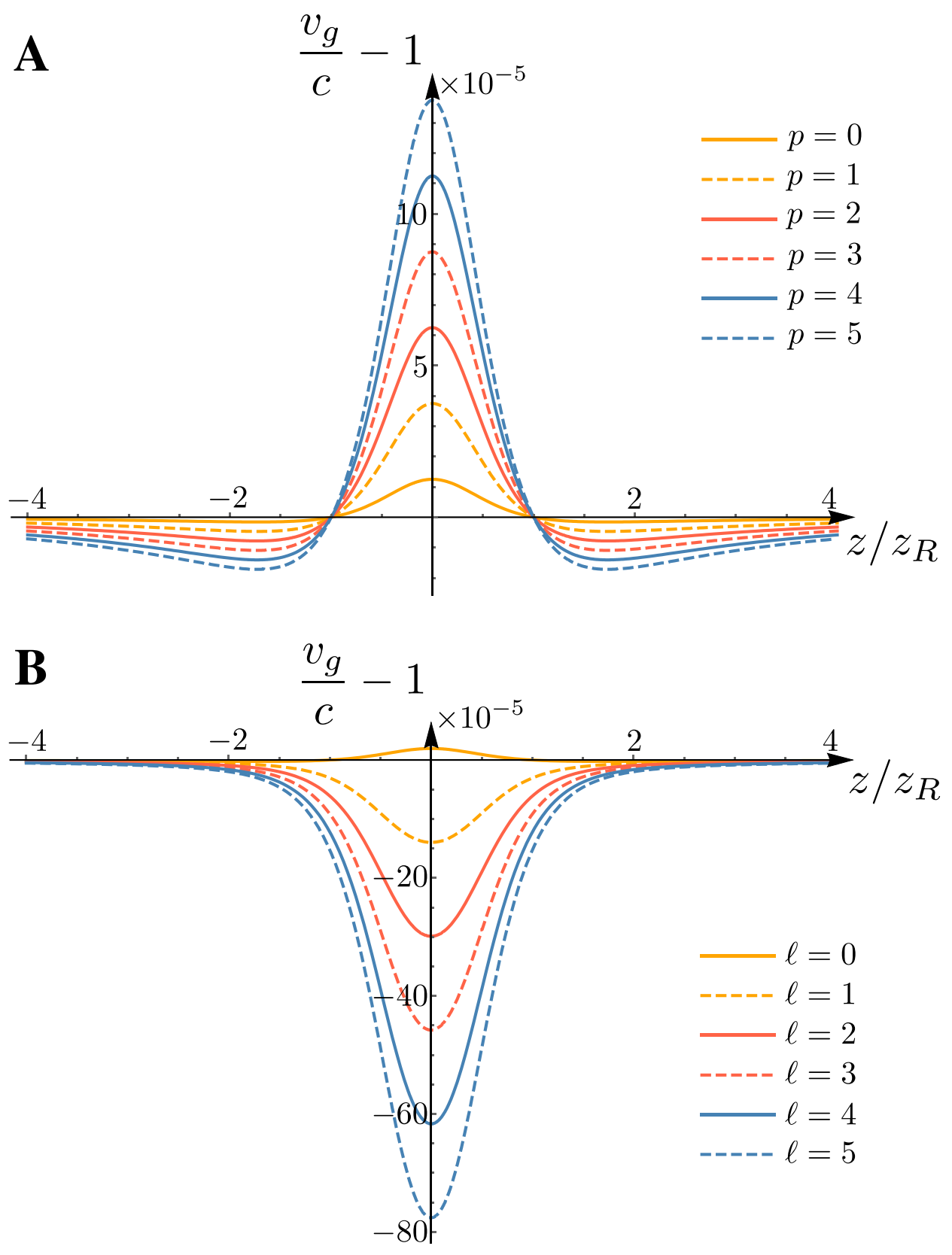

**Fig. 2**. **Group velocities of focused Laguerre-Gauss modes as a function of propagation distance**. **A** Group velocities as a function of propagation distance for LG modes with $\ell = 0$ and various values of $p$. These velocities pertain to the on-axis case, $r = 0$. Three different propagation regions can be identified; a first subluminal zone $z < -z_R$, a superluminal zone $|z| < z_R$, and a second subluminal zone $z > z_R$, where $z_R = \omega\, w_0^2/(2c)$ is the Rayleigh range, at which all modes travel at speed $c$. **B** Propagation dependence of group velocities for LG beams characterized by $p = 0$ and different values of $\ell$. These group velocities are calculated not along the beam axis, but at the radial position corresponding to the beam intensity maximum $r_{\text{max}} = w(z)\sqrt{|\ell|/2}$. $r_{\text{max}}$ defines the circle through which will pass all rays associated with a given LG mode in the raytracing picture. With the exception of the $\ell = 0$ case, all modes are found to exhibit subluminal behaviour throughout propagation. These results are obtained from simulations of light at a wavelength of 795 $n$m with a beam waist of 2.5 mm, focused by a thin lens of focal length 400 mm.

Figure 4-**A** shows normalised power measured as a function of delay stage position using the autocorrelation technique described (see the SM), for HyGG modes with various values of $\ell$. As can be seen, measurable shifts occurred between all HyGG modes presented in the figure. These peak shifts arise from differences in pulse arrival time induced by the subluminal speeds experienced by different HyGG beams during propagation. In particular, the shift in peak position between the $LG_{0,0}$ and $HyGG_6$ modes

is approximately 7 $\mu$m, corresponding to a time delay of 23 fs, from which can be inferred a maximum fractional group velocity drop of 0.1% relative to $c$. This fractional velocity drop is determined from the theoretical curves plotted in Fig. 2. The velocity drop recorded here is that corresponding to the $z = 0$ propagation position, and therefore represents the slowest pulse propagation speed reached during the experiment. Time delays experienced by various $\text{HyGG}_\ell$ modes relative to the $\text{LG}_{0,0}$ reference mode are shown in Fig. 4-**B**. These results reveal the expected linear dependence of pulse arrival time on $\ell$ for a specific propagation distance (see the SM). This dependence may be understood with reference to Eq. (1), which indicates that phase is a function of the index $\ell$ and the radial curvature $R(z, \omega)$.While the $\ell$ index is clearly shown to play an important role in determining the group velocity of an LG mode, the effect of the $p$ index is far less pronounced, for reasons discussed earlier. Thus, HyGG and LG beams will experience similar time delays, up to some correction factor. In the far-field, the group velocity of any LG mode asymptotically approaches $c$, see Fig. 2. Therefore, the time delay between $\text{LG}_{p,\ell}$ and $\text{LG}_{0,0}$ reaches a constant value far from the focus.

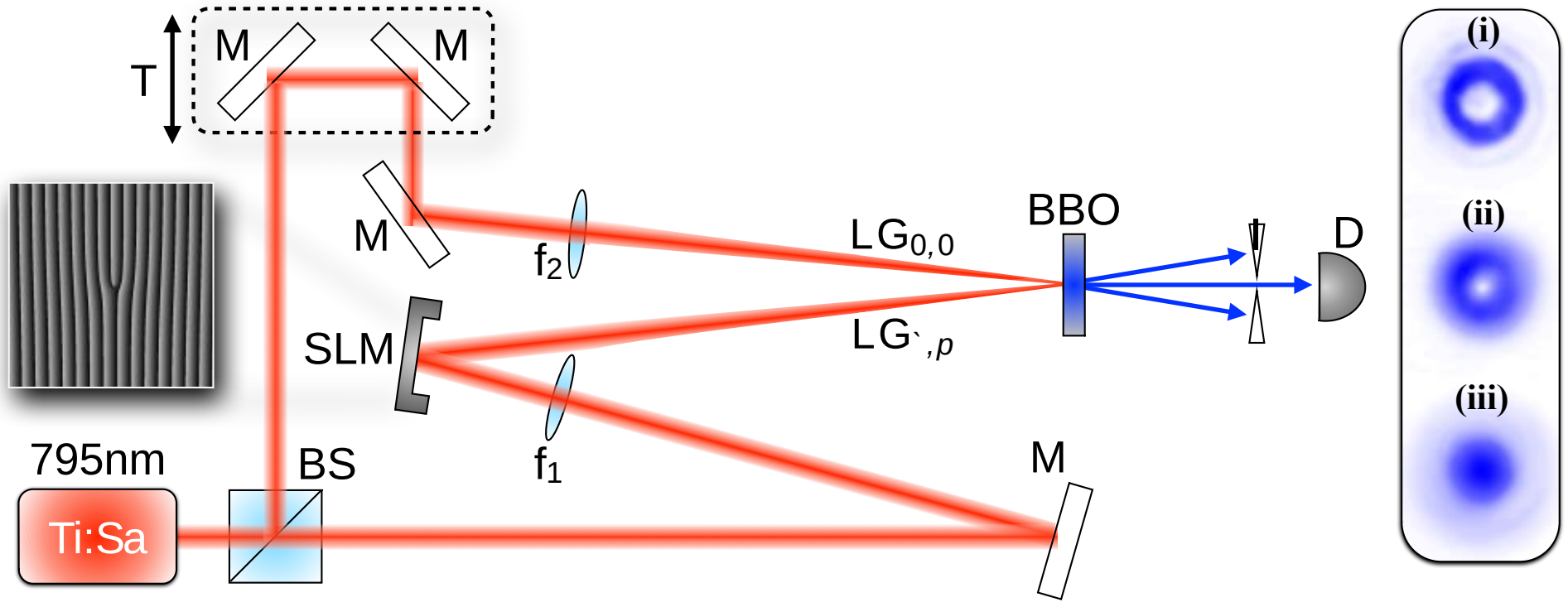

**Fig. 3**. **Experimental setup for measuring subluminal speeds of twisted pulsed beams in vacuum**. The test $\text{HyGG}_\ell$ beam is generated using a spatial light modulator (SLM) onto which is displayed the hologram shown in the figure. Reference $\text{LG}_{0,0}$ and test $\text{HyGG}_\ell$ beams are focused into a type-I $\beta$-barium borate (BBO) crystal. Second-harmonic generation (SHG) output transverse intensity profiles of the test, cross- and reference beams are shown in parts (i)-(iii) of the rightmost inset. When the delay between the test and reference arms is on the order of the pulse duration, one photon from each can be up-converted to produce an SHG photon. By conservation of linear momentum, this photon will exit the BBO crystal at the bisector between the SHG beams produced by the test and reference beams alone. Conservation of OAM further requires that the SHG outputs (i)-(iii) carry OAM values of $2\ell$, $\ell$ and 0 *(22, 23)*. The inset shows beam profiles obtained experimentally for the case $\ell = 1$. Note that the hologram shown here generates a pure $\ell$ mode with an infinite superposition of $p$ modes, where most of the power resides in the fundamental $p = 0$ mode.

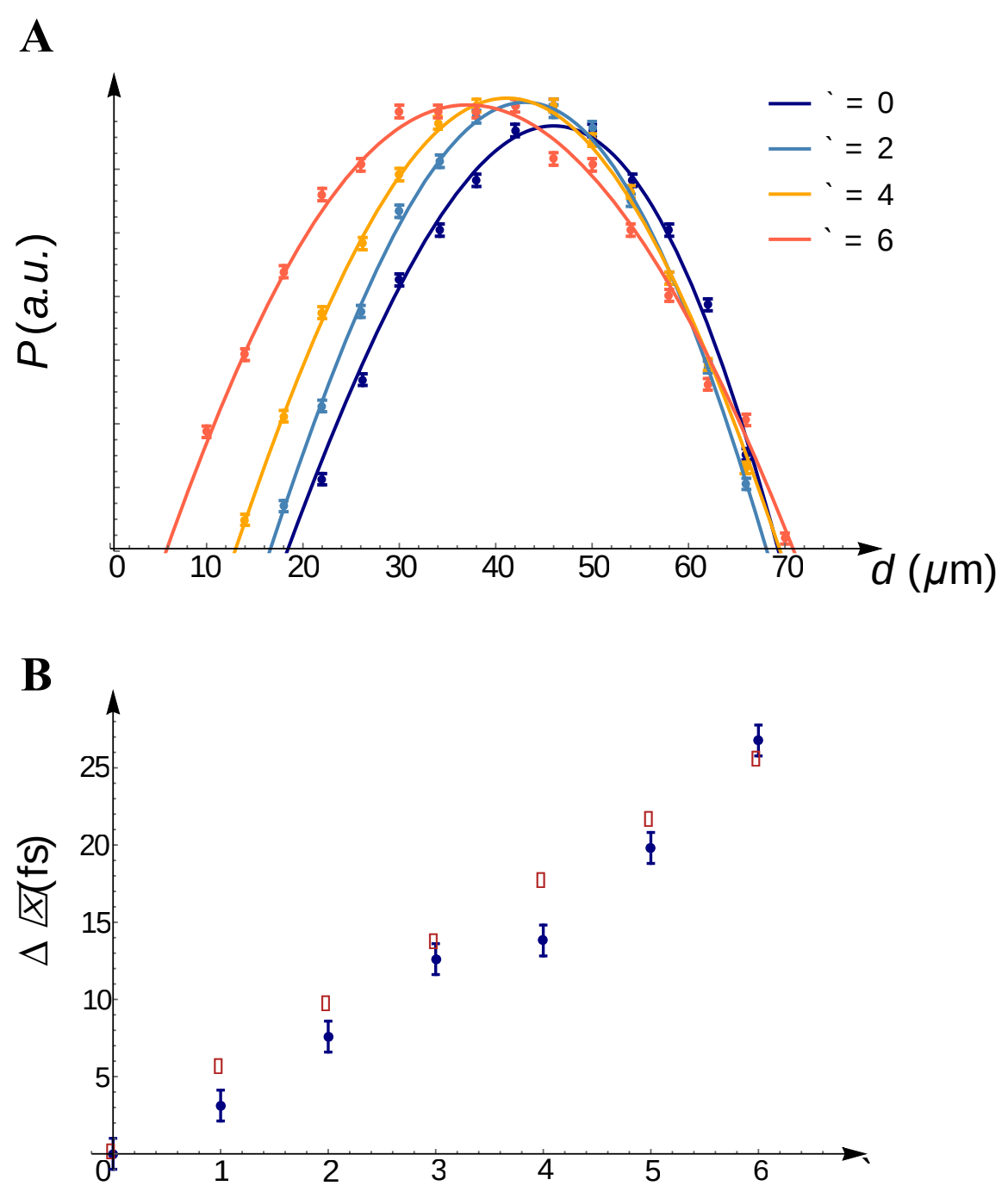

**Fig. 4**. **Subluminal light propagation of HyGG modes in free space**. **A** Normalised power obtained from the auto-correlation technique described in the text as a function of delay stage position, $d$, for HyGG beams with $\ell = 0, 2, 4, 6$. Raw data are presented along with the fits used to estimate pulse peak positions. Power was measured at each stage position over a period of sufficient length to suppress Poissonian noise. The widths of the auto-correlation traces were found to fall in the range of $201 \pm 6\,\mu$m, but in the interest of clarity, only the peaks of the auto-correlation traces are shown. **B** Arrival times of $\text{HyGG}_\ell$ pulses relative to an $\text{LG}_{0,0}$ reference pulse extracted from peak positions determined directly from experimental data. A linear relationship between arrival time and $\ell$ is observed, and the experimental data (blue dots) are plotted along with theoretical data (red dots) obtained from the group velocity expression derived in the SM.

A number of conclusions can be drawn from the investigation presented here. First, the group velocities of twisted LG beams (HyGG beams) have been shown to differ measurably from those of Gaussian pulses, even in free space. This surprising effect can be interpreted as revealing the "pseudo-dispersive" character of the vacuum itself, as diffractive effects in vacuum result in wavelength-dependent phase and group velocities. Further, exotic phase and group velocity effects inevitably arise for any physically realizable beam, for these must necessarily possess nontrivial spatial amplitude and phase structures, which result in other-than-luminal propagation speeds. For the particular case of twisted light, spatial structure can lead to slow light effects whose magnitudes depend directly on various beam parameters, including the modal $\ell$ and $p$ indices.

These findings carry great practical significance, particularly for classical and quantum communication and quantum information with twisted light *(24-26)*. Unless differences induced in LG mode pulse arrival times are compensated for, the sub- and super-luminal effects reported here could result in out-of-sequence detection of pulses, or the failure of quantum logic gates. A photon in an OAM qubit state $|\psi_{\alpha,\beta}\rangle = (\alpha\,|0\rangle + \beta\,|\ell\rangle)$, where $|0\rangle$ and $|\ell\rangle$ respectively represent $\mathrm{LG}_{0,0}$ and $\mathrm{LG}_{0,\ell}$ modes, generated by a sender (Alice) at time, $t = 0$, will travel at a group velocity that depends upon the values of $\alpha$ and $\beta$. As a result, the photon's arrival time at the receiver (Bob) will be given by $t_{\alpha,\beta} = \int_{\mathrm{z=source}}^{\mathrm{z=reciever}} dz/v_{\mathrm{g}}^{\alpha,\beta}(z)$. As this treatment shows, the message will reach Bob at a time that will depend upon $\alpha$ and $\beta$. More generally, photons associated with an arbitrary mode $|\psi_{\{c_\ell\}}\rangle = \sum_{\ell=0}^{+\infty} c_\ell\,|\ell\rangle$ will exhibit arrival times $t_{\{c_\ell\}}$, which will depend upon the set $\{c_\ell\}$ of amplitudes associated with each LG component comprising the beam of interest.

Photon OAM has been shown to represent a valid state label which can be used to distinguish one photon from another. From a quantum information standpoint, therefore, one could imagine how a Hong-Ou-Mandel (HOM)-type experiment could exhibit great sensitivity to the slow-light effects explored here. In particular, a HOM experiment carried out, for example, with one photon in an $\mathrm{LG}_{0,0}$ mode, and the other in a coherent superposition of $\mathrm{LG}_{0,0}$ and $\mathrm{LG}_{0,1}$ modes, would already be expected to register only a limited dip, due to the partial distinguishability of the two photons on the basis of their OAM. However, our work has shown that a second effect will also be at play: in addition to being damped due to the partial distinguishability of the incident photons, one would also expect the HOM dip obtained in such an experiment to be *skewed* as a direct consequence of the different arrival times of the two photons. Given the critical importance of the HOM effect to quantum information protocols, this second effect may have a wide range of practical consequences.

Notably, our results should not be misattributed to lens thickness effects *(27)*, which have been avoided by removing any lenses in the path of the LG beam following its generation. We have shown that time delays on the order of several femtoseconds can arise between LG modes due to the group velocity effects we have explored, depending on the geometries of the sender and receiver optics. As a result, any communication scheme *(28, 29)* or computation protocol *(30)* relying upon twisted light must account for the slow and fast light effects that we have demonstrated.

## Methods

**Experimental setup:** The output of a 100 fs pulsed Ti:Sa laser operating with a repetition rate of 82 MHz and an average power of 300 mW at a central wavelength of 795 $n$m is split into two arms by means of a non-polarising beam splitter (BS). In the test arm, the incident beam is made to pass through a lens placed immediately prior to a kinoform displayed on a Pluto-HOLOEYE spatial light modulator (SLM), which reshapes the spatial distribution of the incoming pulsed beam into a desired $\mathrm{HyGG}_\ell$ mode. In order to generate a pure $\mathrm{LG}_{p,\ell}$ mode, one would have to use an intensity masking technique. The downside to this method is the power reduction in the beam that accompanies the masking strategy, a crucial consideration in a nonlinear optical experiment, as the efficiency of the SHG signal depends on the intensity of the pump beam. No lenses are used after the hologram, in order to avoid the introduction of additional artificial time delays. Since the different LG modes considered have intensity maxima at different radii, a lens placed in the optical path of the beam would introduce an additional time delay, which would depend on the lens thickness at the radius of the beam's intensity maximum. We note that one may be tempted to overcome this difficulty by making use of flat (Fresnel) lenses. However, this would result in a distortion of the pulse shape and spectrum *(27)*.

The first diffracted order obtained from the reflection of the test beam off the SLM is selected using an iris placed at the Fourier plane of the lens. The test and reference pulses are respectively focused by means of $f_1 = 400$ mm and $f_2 = 500$ mm lenses into a $L = 500\ \mu$m thick type-I $\beta$-barium borate (BBO) crystal, where they are spatially overlapped. The Rayleigh range, $z_R = 2$ cm $\approx 40$ L, associated with the test beam is much longer than the crystal length. The cross-SHG signal is isolated from the two other SHG outputs and fundamental beams by means of an iris followed by an interference filter. The power associated with this signal is measured over a period of one second using a Newport power meter with a silicon detector.

**Cross-correlation:** The second-order polarisation arising from the reference ($\mathrm{LG}_{0,0}$)-test ($\mathrm{LG}_{p,\ell}$) beam interaction is given by $P_{0,\ell}^{(2)}(t) = \epsilon_0\, \chi^{(2)} \mathrm{LG}_{0,0}(t)\, \mathrm{LG}_{0,\ell}(t-\tau)$, where $\mathrm{LG}_{0,\ell}(t)$ represents the time-dependent electric field associated with the $\mathrm{LG}_{0,\ell}$ mode, $\tau$ indicates the time delay between the two pulses, and $\chi^{(2)}$ is the second-order nonlinear susceptibility of the BBO crystal. The intensity associated with the cross-beam SHG signal is then proportional to the product $\left|\mathrm{LG}_{0,0}(t)\right|^2 \left|\mathrm{LG}_{0,\ell}(t-\tau)\right|^2$. Hence, maximal cross-beam SHG will be observed when the delay is set to be zero, $\tau = 0$. However, due to the Gaussian temporal profile of the pulses, some non-collinear SHG will result even in the case of small non-zero time delays. As the time delay between both pulses increases, the non-collinear SHG signal decreases gradually, disappearing when $\tau$ significantly exceeds the pulse temporal duration.

**Acknowledgments:** The authors thank Peter Banzer and Miles Padgett for fruitful discussions. F.B. and H.M. acknowledge the support of the Natural Sciences and Engineering Research Council of Canada (NSERC). R.W.B. and E.K. acknowledge the support of the Canada Excellence Research Chairs (CERC) Program. For all data-related enquiries, readers are invited to contact E.K.

**Author Contributions:** E.K.: Conceived the idea; F.B., J.H., H.M. and E.K.: Designed the experiment; F.B., J.H., H.M.: Performed the experiment; F.B., J.H., H.M. and E.K.: Analyzed the data; R.W.B. and E.K.: supervised all aspects of the project. All authors discussed the results and contributed to the text of the manuscript.

**Competing interests:** The authors declare no competing interests. Correspondence and requests for materials should be addressed to E.K. (ekarimi@uottawa.ca).

**Supplementary Materials:**

More details on derivation of phase and group velocity are given in supplementary.

Figure S1. Time delay between LG mode pulses and a plane wave pulse as a function of position along the beam axis.

References (*1, 31*)

14 December **2014**

# Supplementary Materials for Observation of subluminal twisted light in vacuum

**Authors:** Frédéric Bouchard[1,2], Jérémie Harris[1,2], Harjaspreet Mand[1,2], Robert W. Boyd[1,2,3], Ebrahim Karimi[1,2]*

[1]Department of Physics, University of Ottawa, 25 Templeton St., Ottawa, Ontario, K1N 6N5 Canada.

[2]The Max Planck Centre for Extreme and Quantum Photonics, University of Ottawa, Ottawa, Ontario, K1N 6N5, Canada.

[3]Institute of Optics, University of Rochester, Rochester, New York, 14627, USA.

*Correspondence to: Ebrahim Karimi (ekarimi@uottawa.ca).

## Phase and Group Velocities

A surface of constant phase, also known as a cophasal surface, associated with a monochromatic electric field $E(\mathbf{r},t) = A(\mathbf{r})e^{i(\omega\, t-\Phi(\mathbf{r}))}$, where $A(\mathbf{r})$ and $\Phi(\mathbf{r})$ respectively denote the amplitude and phase of the electric field, is given by *(1)*

$$\omega\, t - \Phi(\mathbf{r}) = \text{constant}, \qquad \text{(S1)}$$

whence

$$\delta\big(\omega\, t - \Phi(\mathbf{r})\big) = 0. \qquad \text{(S2)}$$

The phase velocity describes the rate at which a point of constant phase travels through space, i.e. $v_{\text{ph}} = \delta r_{\text{ph}}/\delta t$, and can be calculated by

$$\omega\, \delta t - \nabla\Phi(\mathbf{r})\cdot\delta\mathbf{r} = 0. \qquad \text{(S3)}$$

Equation (S3) can be rearranged by introducing a unit vector $\hat{\mathbf{r}} = \delta\mathbf{r}/\delta r_{\text{ph}}$, yielding

$$v_{\text{ph}} = \frac{\delta r_{\text{ph}}}{\delta t} = \frac{\omega}{\mathbf{r}\cdot\nabla\Phi(\mathbf{r})} = \frac{\omega}{|\nabla\Phi(\mathbf{r})|}, \qquad \text{(S4)}$$

where we have used the fact that $\hat{\mathbf{r}} = \nabla\Phi(\mathbf{r})/|\nabla\Phi(\mathbf{r})|$ for propagation of a wavefront.

For polychromatic fields, which in general may be expressed in the form

$$E(\mathbf{r},t) = \int_{-\infty}^{\infty} \tilde{A}(\mathbf{r},\omega)e^{i(\omega\, t-\Phi(\mathbf{r},\omega))}d\omega,$$

a group velocity may be assigned to the pulse envelope. Here, $\tilde{A}(\mathbf{r},\omega)$ is the angular frequency amplitude and represents the amplitude of the field in frequency space. The group velocity may be derived by expressing the electric field as a product of its carrier and envelope components, such that $E(\mathbf{r},t) = A(\mathbf{r},t)e^{i(\omega_0\, t-\Phi(\mathbf{r},\omega_0))}$, where $\omega_0$ stands for the central angular frequency of the field. From these two expressions for the electric field, it can be seen that

$$A(\mathbf{r},t) = \int_{-\infty}^{\infty} \tilde{A}(\mathbf{r},\omega)e^{i\{(\omega-\omega_0)t-(\Phi(\mathbf{r},\omega)-\Phi(\mathbf{r},\omega_0))\}}d\omega. \qquad \text{(S5)}$$

The phase $\Phi_{\text{envelope}}$ of the field envelope therefore evolves according to $\Phi_{\text{envelope}} = (\omega - \omega_0)t - (\Phi(\mathbf{r},\omega) - \Phi(\mathbf{r},\omega_0))$. When $\Phi(\mathbf{r},\omega)$ is Taylor-expanded to first order in frequency around $\omega_0$, i.e. $\Phi(\mathbf{r},\omega) \simeq \Phi(\mathbf{r},\omega_0) + (\omega - \omega_0)\partial_{\omega_0}\Phi(\mathbf{r},\omega)$, the condition for constant envelope phase is found to be

$$\delta\left(\Phi_{\text{envelope}}\right) = 0; \quad \text{(S6)}$$

therefore, we have

$$(\omega - \omega_0)\delta\left(t - \partial_{\omega_0}\Phi(\mathbf{r},\omega)\right) = (\omega - \omega_0)\left(\delta t - \partial_{\omega_0}\nabla\Phi(\mathbf{r},\omega)\cdot\delta\mathbf{r}\right) = 0 \quad \text{(S7)}$$

Defining a unit vector for the position of the envelope, $\hat{\mathbf{r}} = \delta\mathbf{r}/\delta r_{\mathrm{g}}$, then gives

$$\begin{aligned} v_{\mathrm{g}} &= \frac{\delta r_{\mathrm{g}}}{\delta t} \\ &= \frac{1}{\hat{\mathbf{r}}\cdot\nabla\partial_{\omega_0}\Phi(\mathbf{r},\omega)} = \frac{1}{\left|\nabla\partial_{\omega_0}\Phi(\mathbf{r},\omega)\right|}. \end{aligned} \quad \text{(S8)}$$

The amplitude of a Laguerre-Gauss beam is given by

$$\mathrm{LG}_{p,\ell}(r,\varphi,z) = \frac{C_{p,\ell}}{w(z)}\left(\frac{r\sqrt{2}}{w(z)}\right)^{|\ell|}\exp\left(-\frac{r^2}{w^2(z)}\right)L_p^{|\ell|}\left(\frac{2r^2}{w^2(z)}\right)e^{i\Phi^{\mathrm{LG}}}, \quad \text{(S9)}$$

where $p$ is the radial index (positive integer), $\ell$ is the azimuthal index (integer), $r$ and $\varphi$ are the transverse cylindrical coordinates, $C_{p,\ell}$ are normalization constants, $w(z) = w_0(1 + (z/z_R(\omega))^2)^{1/2}$ is the beam radius, $L_p^{|\ell|}(x)$ are the generalized Laguerre polynomials, and $\Phi^{\mathrm{LG}}$ contains all of the phase terms of the Laguerre-Gauss beam. The LG beam phase may be expressed explicitly as

$$\Phi_{p,\ell}^{\mathrm{LG}}(r,\varphi,z,\omega) = \frac{\omega}{c}z + \frac{\omega}{c}\frac{r^2}{2R(z,\omega)} + \ell\varphi - (2p + |\ell| + 1)\zeta, \quad \text{(S10)}$$

where $\zeta \coloneqq \arctan\left(\frac{z}{z_R(\omega)}\right)$ is the Gouy phase. The radius of curvature $R(z,\omega)$ is obtained from the ABCD law for the case of a Gaussian beam incident on a thin lens at its waist, and subsequently focused to a new waist, and is given by *(31)*

$$R(z,\omega) = \frac{\left(\frac{z+d(\omega)}{z_{R,0}(\omega)}\right)^2 + \left(1 - \frac{z+d(\omega)}{f}\right)^2}{\left(\frac{z+d(\omega)}{z_{R,0}^2(\omega)}\right) - \frac{1}{f}\left(1 - \frac{z+d(\omega)}{f}\right)}, \quad \text{(S11)}$$

where $d(\omega) = f\left(1 + \left(f/z_{R,0}(\omega)\right)^2\right)^{-1}$ is the distance between the lens and the focus, $f$ is the focal length of the lens, and $z_{R,0}(\omega)$ is the Rayleigh range of the beam prior to the lens. From the expression (S10) for the phase, one can obtain an explicit form for the group velocity of an LG beam by direct application of Eq. (S8),

$$v_{\mathrm{g}}(\mathbf{r},\omega) = \frac{1}{\left|\nabla\partial_{\omega}\Phi_{p,\ell}^{\mathrm{LG}}(\mathbf{r},\omega)\right|}. \quad \text{(S12)}$$

This was the approach taken to produce the theoretical curves shown in Fig. 2 for LG modes having different indices $p$ and $\ell$. The group velocity associated with a pulse exhibits a dependence on $\mathbf{r}$. The radii at which velocity measurements were made correspond to the coordinates through which the classical rays associated with each beam will pass, according to the ray tracing picture. Measurements plotted in Fig. 4 naturally reflect the group velocity of LG modes at $r_{\text{max}} = w(z)\sqrt{|\ell|/2}$ for $p = 0$ modes having different values of $\ell$.

We note also that two LG beams characterized by equal central wavelengths and $r_{\text{max}}$ values, but different indices $\ell$ and, correspondingly, different beam waists, will propagate at different group velocities. Even if the classical ray trajectories associated with two LG modes are made to coincide at the focus by changing the beam waist, their Rayleigh ranges will differ, and scale inversely with $\ell$. Therefore, it is impossible to compensate for this effect by simply changing the beam waist; it is a beam property rather than a property of the optics. It should also be noted that the results shown in Fig. 2 are wavelength and beam waist dependent.

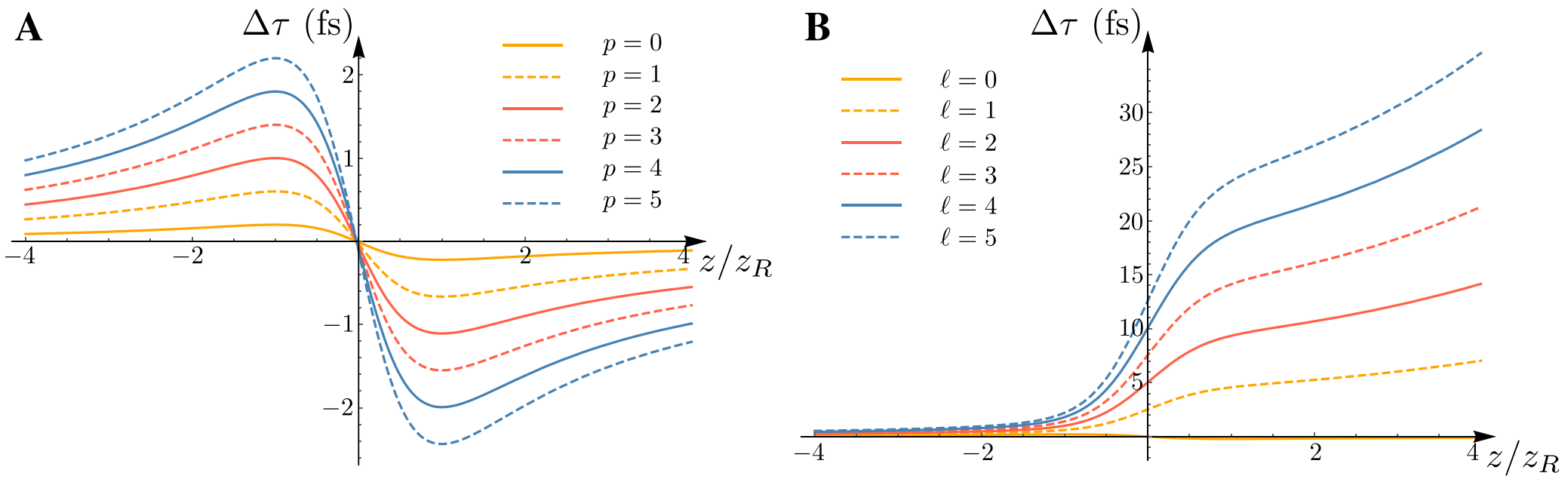

Fig. S1: **Time delay between LG mode pulses and a plane wave pulse as a function of position along the beam axis. A** Theoretical arrival times of $\text{LG}_{p,0}$ pulses relative to a plane wave pulse travelling at $c$, as a function of position along the beam axis. Since the maximal intensity of $\text{LG}_{p,0}$ modes is found on-axis (at $r = 0$), these relative arrival times do not include any contribution from the phase-front radius of curvature. As a result, any deviations from exactly luminal propagation may be attributed entirely to changes in the $p$ index itself. **B** Theoretical plot showing arrival times of $\text{LG}_{0,\ell}$ pulses as a function of propagation distance, relative to a plane wave pulse travelling at $c$. The time delays between LG pulses with nonzero $\ell$ indices and the plane wave pulse are monotonically-increasing functions of propagation distance for $\ell \neq 0$. This is due to the fact that LG beams with nonzero $\ell$ are characterized by subluminal group velocities throughout propagation. In each case, the plotted time delays are calculated at the radius of maximum intensity, $r_{\text{max}}(z)$, and therefore include the contribution of the phase-front radius of curvature to the group velocity of the LG pulse.

## Hypergeometric-Gaussian modes

A given Hypergeometric-Gaussian mode with well-defined $\ell$ can be regarded as an infinite superposition of $\text{LG}_{p,\ell}$ modes with fixed $\ell$ and different $p$ indices. In a general, a HyGG beam generated from a pitch-fork hologram finds most of its power peaked in the $\text{LG}_{0,\ell}$ mode. Perhaps the most important correction that one needs to include while dealing with HyGG modes is the radius of maximum intensity $r_{\max}$. In the following, we carry out a calculation to obtain an expression for the $r_{\max}$ of a HyGG beam at the focus.

The HyGG modes (in dimensionless coordinates) can be expressed as a superposition of modified Bessel function of the first kind.

$$u_{-|m|,m}(\rho) \propto e^{\frac{1}{2}\left(\frac{-2i}{1+\zeta}+\frac{1}{i\zeta+\zeta^2}\right)\rho^2}\rho\left[I_{\frac{|m|-1}{2}}\left(\frac{\rho^2}{2\zeta(1+i\zeta)}\right)-I_{\frac{|m|+1}{2}}\left(\frac{\rho^2}{2\zeta(1+i\zeta)}\right)\right],$$

where $\rho = r/w_0$ and $\zeta = z/z_{\mathrm{R}}$ are the dimensionless radial and longitudinal coordinates, respectively. We consider here only the radial dependence of the mode (for reasons that will be obvious as we proceed). The squared modulus of the modal amplitude is then calculated,

$$\left|u_{-|m|,m}(\rho)\right|^2 \propto e^{\left(\frac{\rho^2}{1+\zeta^2}\right)}\rho^2\left|I_{\frac{|m|-1}{2}}\left(\frac{\rho^2}{2\zeta(1+i\zeta)}\right)-I_{\frac{|m|+1}{2}}\left(\frac{\rho^2}{2\zeta(1+i\zeta)}\right)\right|^2.$$

At this stage in the derivation, we assume that the term in the argument of the Bessel functions is very small. This condition is satisfied as the beam propagates away from the near field, at larger values of z. We can then approximate the modified Bessel function by,

$$I_\alpha(z) \approx \frac{1}{\Gamma(\alpha+1)}\left(\frac{z}{2}\right)^\alpha.$$

This approximation is valid in the limit where

$$0 < \left|\frac{\rho^2}{2\zeta(1+i\zeta)}\right| \ll \sqrt{\frac{|m|+1}{2}}.$$

The intensity of the beam can then be expressed in the following form,

$$\left|u_{-|m|,m}(\rho)\right|^2 \propto e^{\left(\frac{\rho^2}{1+\zeta^2}\right)}\rho^2\frac{1}{\Gamma\left(\frac{|m|+1}{2}\right)}\left(\frac{\rho^2}{4\zeta\sqrt{1+\zeta^2}}\right)^{|m|-1}\left[1-\frac{1}{|m|+1}\left(\frac{\rho^2}{1+\zeta^2}\right)+\left(\frac{1}{|m|+1}\right)^2\frac{\rho^4}{4(\zeta^2+\zeta^4)}\right].$$

We proceed with an additional approximation, by neglecting the last term in the expression above. This is physically reasonable since z is assumed to be large in this case. Hence,

$$\left|u_{-|m|,m}(\rho)\right|^2 \propto e^{\left(\frac{\rho^2}{1+\zeta^2}\right)}\rho^2\frac{1}{\Gamma\left(\frac{|m|+1}{2}\right)}\left(\frac{\rho^2}{4\zeta\sqrt{1+\zeta^2}}\right)^{|m|-1}\left[1-\frac{1}{|m|+1}\left(\frac{\rho^2}{1+\zeta^2}\right)\right].$$

If we concern ourselves only with the radial dependencies,

$$\left|u_{-|m|,m}(\rho)\right|^2 \propto e^{\left(\frac{\rho^2}{1+\zeta^2}\right)}\rho^{2|m|}\left[1-\frac{1}{|m|+1}\left(\frac{\rho^2}{1+\zeta^2}\right)\right].$$

We can now differentiate the expression in order to obtain the maximal value for $\rho$,

$$\partial_\rho\left|u_{-|m|,m}(\rho_{\max})\right|^2 = 0,$$

$$\rho_{\max} = |m|^{1/4}(|m|+1)^{1/4}\sqrt{1+\zeta^2}.$$

Thus we obtain an expression slightly different from the case of LG beams, for which

$$r_{\max}^{(\mathrm{HyGG})}(\ell, z) = w(z)\big(|\ell|(|\ell| + 1)\big)^{1/4}.$$